\documentstyle[psfig]{mn}

\newcommand{\pr}{\prime}
\newcommand{\be}{\begin{equation}}
\newcommand{\ee}{\end{equation}}
\newcommand{\bgrad}{{\bmath\nabla}}
\newcommand{\bea}{\begin{eqnarray}}
\newcommand{\eea}{\end{eqnarray}}
\newcommand{\p}{\prime}

\title[Strong magnetic fields in accretion discs]
      {Stability of accretion discs threaded by a strong magnetic field}

\author[R. Stehle, H.C. Spruit]
{R. Stehle$^{1,2}$
H.C. Spruit,$^{1,3}$\\
$^1$Max Planck Institute for Astrophysics, Box 1523, D-85740 Garching, Germany
\\           
$^2$Astronomy Group, University of Leicester, Leicester, LE1 7RH, UK
\\
$^3$Astronomical Institute `Anton Pannekoek', Kruislaan 403, 1098 SJ Amsterdam,
Netherlands  }
\date{}

\pagerange{\pageref{firstpage}--\pageref{lastpage}}
\pubyear{2001}

\date{submitted: 1998 February 25, accepted 2000 November 7}

\begin{document}

\maketitle

\label{firstpage}

\begin{abstract}
We study the stability of poloidal magnetic fields
anchored in a thin accretion disc. The two-dimensional  
hydrodynamics in the disc plane is followed by a grid-based
numerical simulation including the vertically integrated magnetic 
forces. 
The 3--dimensional magnetic field outside the disc is calculated
in a potential field approximation from the magnetic 
flux density distribution in the disc. 
For uniformly rotating discs we confirm numerically 
the existence of the interchange instability as predicted 
by Spruit, Stehle \& Papaloizou (1995).
In agreement with predictions from the shearing sheet model, 
discs with Keplerian rotation are found to be stabilized by the
shear, as long as the contribution of magnetic forces to 
support against gravity is small. 
When this support becomes significant, we find a global 
instability which transports angular momentum outward 
and allows mass to accrete inward. 
The instability takes the form of a $m=1$ rotating 
`crescent', reminiscent of the purely hydrodynamic 
nonlinear instability previously found in pressure-supported
discs.
A model where the initial surface mass density 
$\Sigma(r)$ and $B_{\mathrm{z}}(r)$ decrease with 
radius as power laws shows transient mass accretion 
during about 6 orbital periods, and settles into a 
state with surface density and field strength  
decreasing approximately exponentially with radius. 
We argue that this instability is likely to be 
the main angular momentum transport mechanism in
discs with a poloidal magnetic field sufficiently strong
to suppress magnetic turbulence. It may be especially relevant
in jet-producing discs.
\end{abstract}


\section{Introduction}

In the formation and evolution of Young Stellar Objects 
magnetic fields are thought to play a key role. 
Magnetic fields, beside rotation and thermal pressure, stabilize 
molecular clouds from gravitational infall.
By magnetic braking, thermal cooling or diffusive magnetic 
processes like ambipolar diffusion a supercritical 
cloud can form. 
The subsequent collapse, which proceeds preferentially 
along the rotation axis and the magnetic field lines, 
results in a dense central object, i.e. a protostar 
and an accretion disc (Mestel 1965;  Shu et al. 1993;
see Tomisaka 1995 for a numerical study).
After the first, dynamical collapse we expect the 
disc around the central object to be still threaded 
by a fraction of the interstellar magnetic flux of 
the original cloud.

Though microscopic diffusion processes are important 
for the evolution of magnetic fields in star forming 
regions, effective diffusion of matter across field lines
is also possible through instabilities driven by the 
magnetic field itself. 
In the molecular cloud cores where the magnetic field 
is thought to disengage itself from the matter, the
gravitational and magnetic energy densities are believed 
to be of comparable magnitude 
(see McKee et al. 1993 for a review). 
This is just the condition under which magnetic instabilities 
can operate at rates competitive with the gravitational 
collapse rate. 
It would be somewhat of a coincidence if microscopic 
diffusion were always the dominant process, and effective
diffusion by magnetically-driven instabilities never played a role 
in the entire contraction process from cloud to star. 
This is one important reason to study the possible effects 
of non--axisymmetric magnetic instabilities in the process 
of star formation.

Another reason is provided by the possible connection of 
magnetic fields in discs with outflows and jets. 
A remnant of the cloud core's magnetic flux, anchored 
in the accretion disc, represents a poloidal magnetic 
field component with the right geometry to launch and 
accelerate disc gas along the magnetic field lines 
(Bisnovatyi-Kogan and Ruzmaikin 1976; Blandford \& Payne 1982).
P Cygni line profiles in the spectra of Young Stellar Objects
might find their explanation in such disc winds 
(Edwards, Ray \& Mundt 1993).

In addition to their role in accelerating outflows, poloidal 
magnetic fields anchored in the accretion disc may also
play an important role in the collimation of bipolar outflows and jets 
(Blandford 1993; Spruit, Fogglizio \& Stehle 1997).
We refer to Hughes (1991) for an extensive 
introduction into the observations and physics 
of beams and jets. For a review and tutorial introduction to 
the magnetic acceleration model see 
Spruit 1996; for some issues of current interest see Ogilvie 
and Livio (1998), Cao and Spruit (1994, 2000).

The ability of the disc to produce a magnetically accelerated 
outflow depends rather critically on the strength and radial 
distribution of the poloidal field at its surface.
This distribution, in turn, is determined by the rate at 
which an advected large scale magnetic field  in the disc 
is able to diffuse outward against the inward drift velocity 
compressing it.

Van Ballegooijen (1989) studied the radial 
transport of magnetic field lines assuming an isotropic 
turbulent viscosity $\nu$, related to an equally isotropic 
magnetic diffusivity $\eta$ by a constant ratio, the
magnetic Prandtl number. 
Along the same lines Lubow, Papaloizou \& Pringle (1994)
studied the magnetic field dragging by turbulent processes 
in an accretion disc which was initially threaded by an 
externally imposed magnetic field. 
The conclusion from these studies is that inward 
dragging of an external field is difficult to achieve, 
if the assumption of an isotropic magnetic Prandtl number
of order unity holds. 
It follows that one should expect to see, in this case, 
only internally generated fields
like those obtained in numerical simulations 
(Stone et al. 1996; Brandenburg et al. 1995). 

If the initial magnetic field is sufficiently strong to 
contribute to support against gravity, however, 
as appears to be the case in cloud cores, it is likely to 
suppress these dynamo processes, since a weak-field 
instability (Balbus \& Hawley 1992) is thought to be 
an essential ingredient in this kind of dynamo. 
In this case, which we shall call here the strong field 
case, the most plausible source of turbulence in the disc 
are non--axisymmetric instabilities caused by the strong magnetic 
field itself.

One well-known instability is {\it interchange}, 
a local instability driven by the magnetic field energy, and
which plays an important role in controlled fusion 
devices and in solar magnetic fields. 
The example of solar prominences shows that these 
instabilities can be quite efficient in transporting 
mass across field lines (cf. Priest 1982). 
Linear interchange instability
has been studied previously by Spruit \& Taam (1990) 
for uniformly rotating discs, and for differentially 
rotating discs by Spruit, Stehle \& Papaloizou (1996) 
and Lubow \& Spruit (1996), in a shearing sheet 
approximation. The instability appears when the ratio
$B_z/\Sigma$ of the vertical magnetic field strength 
to the surface mass density decreases with distance 
from the centre of the disc. Its behavior is similar to 
convection, with the gradient of $B_z/\Sigma$ replacing
the entropy gradient.

As in the case of solar prominences, it is the magnetic tension 
force due to the curvature of the field lines in the $r-z$-plane 
that drives the instability in geometrically thin discs 
(Anzer 1967; Spruit and Taam 1992; 
for a detailed analysis of the magnetic forces in thin discs
see Ogilvie 1997). Since non--axisymmetry is also crucial, this  
makes the magnetic field three-dimensional.  Our study thus
differs conceptually from the numerical study by Kaisig Tajima 
and Lovelace (1992), where a 2-dimensional cylindrical configuration 
with vertical magnetic field lines was assumed. In that case, the only 
magnetic force is the magnetic pressure gradient. 

A full three-dimensional numerical treatment of the 
problem is made difficult by the very large range 
of characteristic speeds expected in the problem. 
Inside the disc, the Alvf\'en speed is not larger 
than the sound speed, but in the low density 
regions outside the disc it can easily approach 
the speed of light. 
We show here how this difficulty can be 
turned into an advantage, such that only a
part of the problem needs to be done in 3 
dimensions, while the numerically most demanding 
parts can be done using only the two dimensions 
in the disc plane.

In the above, we have introduced the magnetic disc model 
with the example of protostellar discs. 
The physics addressed by our calculations, however, 
is equally applicable to discs in compact binaries or 
active galactic nuclei.

\subsection{poloidal fields vs. dynamo-generated fields}

The magnetic problem we study is in several ways 
complementary to that of the simulations 
done by Stone et al. (1996) or Brandenburg et al. (1995), and Hawley (2000). 
There the magnetic fields are generated locally by a
dynamo process, which can already start from a very weak intitial field. If a strong field is present initially, such that the magnetic energy density is comparable to the thermal energy density, the dynamo process is suppressed. This limit can be written in the form
\be
{B^2\over 4\pi\Sigma}\la c_{\rm s}\Omega, \label{blim}
\ee
where $\Sigma$ is the surface mass density and $c_{\rm s}$ the sound speed.

A small scale dynamo process like the magnetic turbulence seen in these simulations does not create a net flux of field lines through the disc (Hawley 2000), and the overall 
field structure is therefore at least of quadrupole order at large distance from the disc.
For advected magnetic field lines, as they are advected radially and anchored in the disc, 
the total magnetic flux through the disc is non--zero. The global structure of the advected magnetic field, as seen from large distances, is thus close to a 
dipole magnetic field distribution. The radial force exerted by such a field is predominantly the tension force, whose magnitude integrated over the disc thickness is $ B_rB_z/2\pi$. A poloidal field can in principle exist up to strengths such that this radial force starts contributing significantly to the support against gravity. This limit can be written as
\be
{B^2\over 4\pi\Sigma}\la \Omega^2r.
\ee
This limit on $B^2$ is a factor $\Omega r/c_{\rm s}$ larger than the strength (\ref{blim}) at which dynamo-generated fields are suppressed. Once strong poloidal fields exist in a disc, they suppress the magnetic turbulence that would make them diffuse out of the disc by van Ballegooijen's argument. There is thus a large range in parameter space where a poloidal field in a disc would be affected only by its own internal instabilities. These are the subject of the present investigation.

\section{Two-dimensional discs with three-dimensional magnetospheres}

We neglect the self--gravity of the disc so
that the central star is the only cause of gravitational 
force acting on the disc. 
Following Spruit \& Taam (1990, hereafter ST, see also
Tagger et al. 1990 for a similar discussion) 
we assume a geometrically thin disc with an internal 
velocity field that does not depend on the vertical 
coordinate (perpendicular to the disc). 
The equations of motion can then be integrated across 
the disc, resulting in a two-dimensional problem 
confined to the plane of the disc. 
This plane can in principle be of arbitrary time-dependent 
shape, as in ST, but we limit the calculations here to 
the case of a plane disc, without displacements of the 
disc surface in the vertical direction 
(i.e. without corrugations or `bending modes'. For results on such
modes, in the same approximation for the magnetic field, see Agapitou 
et al. 1997). 

One may wonder when the approximation of height-independent
velocity fields in the disc is justified. 
The assumption clearly eliminates the possibility
of dynamo-generation of magnetic fields, but is 
appropriate for {\it strong fields} in the sense 
discussed in the preceding section, namely fields 
which contribute, more than the pressure force, 
to support against gravity. 
For such strong fields, winding-up of field lines 
inside the disc can be ignored, because the small 
amount of differential rotation encountered by the 
field line as it crosses from one side of the disc 
to the other is not enough to bend field lines 
significantly. 
Instead, the differential rotation across the disc 
will adjust, due to the magnetic forces, such that 
the rotation rate is constant along a field line. 
In the thin-disc limit, the amount of differential rotation
across the disc vanishes as $H/r$, justifying 
the assumption made (for a discussion of the thin 
limit of magnetic discs, from a more mathematical 
point of view, see Ogilvie 1997). 

Because of the thin disc limit taken, the dominant 
magnetic force in the problem is the curvature force 
due to the bend of the field lines crossing the disc 
(see ST). 
Outside the disc, the magnetic forces dominate over 
fluid forces on account of the low gas density. 
We call this region of low plasma-$\beta$ the 
`magnetosphere' of the disc (not to be confused 
with the magnetosphere of an accreting magnetic star). 
We simplify the physics by assuming the field in 
this magnetosphere to be {\it current free} so that 
it is derivable from a magnetic potential $\psi$. 
This is justified if the matter density in the magnetosphere
is sufficiently low. The magnetospheric field in then approxmately
{\it force free}. The case of the solar corona shows that such a field
is in practice also close to a potential (current free) field. This is
due to the fact that at low $\beta$,  `forced' processes (e.g. 
Parker 1979, for recent numerical simulations see Galsgaard and Nordlund 1997)
are fast, and keep the degree of twisting in the field low. 

In this approximation,  all currents are confined to the plane of the 
disc, and are proportional to the 
jump in the tangential field components across the disc. 
These tangential components ($B_\phi,B_r$) are determined by the normal 
field component through the solution of the (3-dimensional) 
potential problem in the magnetosphere, for which the normal 
component $B_z$ provides the boundary condition.

The magnetic forces acting on the fluid in the disc plane 
are given by the difference in the magnetic stress  
acting on the upper and lower surfaces of the disc. 
They are proportional to the product of the tangential 
and normal components of the magnetic field at the disc 
surface (see ST for details).

In the computations, closed inner and outer boundaries 
are used for the disc ($v_r=0$, so that no mass or magnetic flux 
crosses these boundaries. 
Thus the total magnetic flux through the disc and the 
total disc mass are constant in time.

Apart from the addition of the magnetic force term, 
the hydrodynamical problem is the same as in ordinary 
two-dimensional disc hydrodynamics. 
We use an Eulerian grid with the van Leer (1977) 
scheme for upwind differencing. 
An outline of the model and its basic 
assumptions has been given previously in Stehle (1997).

\section{Equations}
\label{eqs}
To describe the non--axisymmetric disc response resulting from 
large scale magnetic fields, we adopt a thin--disc approximation
in which the vertical velocity vanishes.  We use a cylindrical coordinate system 
($r, \phi, z$) defined with the origin at the central mass $M$.
The mass surface--density is
$\Sigma=\int_{-\infty}^{+\infty} \rho {\mathrm{d}} z$ 
with $\rho=\rho(r,\phi,z)$ the gas density. 

In the thin disc limit $H/r\ll 1$ where $H$ is the disc thickness the only contribution of the Lorentz force per unit surface area is due the magnetic tension of the field lines (ST). This is because for magnetic fields varying on a length scale $L\gg H$, the curvature force is larger than the magnetic pressure gradient by a factor $L/H$.
Then the radial component of the equation of motion 
for the disc, assumed to be inviscid, reads
\begin{eqnarray}
\label{euler_1}
\frac{ \partial v_{\mathrm{r}}}{\partial t} 
&+&
v_{\mathrm{r}} \frac{\partial v_{\mathrm{r}}}{\partial r}
+
\frac{v_{\phi}}{r} \frac{\partial v_{\mathrm{r}}}{\partial \phi}
-
\frac{v_{\phi}^{2}}{r}
\\ \nonumber 
&=&
- \frac{1}{\Sigma} \frac{\partial P}{\partial r}
+
\frac{ B_{\mathrm{z}} [B_{\mathrm{r}}]}{4 \pi \Sigma}
- g,
\end{eqnarray}
where $v_{\mathrm{r}}$ and $v_{\mathrm{\phi}}= r \Omega$ are the
radial and azimuthal components of the velocity vector and $g$is the 
acceleration of gravity due to the central star.
$B_{\mathrm{z}}$ is the vertical magnetic field component at disc midplane and 
$[B_{\mathrm{r}}]=B_{\mathrm{r}}^{+}-B_{\mathrm{r}}^{-}=2B_{\mathrm{r}}^{+}$
the jump of the field vector from 
above ($B_{\mathrm{r}}^{+}$) to below ($B_{\mathrm{r}}^{-}$) the
disc plane. 
Neglecting any possible disc warps it is assumed that the 
field geometry is antisymmetric with respect to the disc plane
(i.e. $B_{\mathrm{z}}^{+}=B_{\mathrm{z}}^{-}$ and 
      $B_{\mathrm{r}}^{+}=-B_{\mathrm{r}}^{-}$, 
  see ST). 
For use in what follows we define the magnetic acceleration in radial direction,
\be g_{\mathrm{m}}=B_{\mathrm{z}}B_{\mathrm{r}}^{+} / 2\pi\Sigma.\ee



The azimuthal equation of motion is
\begin{eqnarray}
\label{euler_2}
\frac{\partial v_{\phi}}{\partial t} 
&+&
v_{\mathrm{r}} \frac{\partial v_{\phi}}{\partial r}
+
\frac{v_{\phi}}{r} \frac{\partial v_{\phi}}{\partial \phi}
+
\frac{v_{\mathrm{r}} v_{\phi}}{r}
\\ \nonumber
&=&
- 
\frac{1}{\Sigma} \frac{\partial P}{r \partial \phi}
+
\frac{ B_{\mathrm{z}} [B_{\phi}]}{4 \pi \Sigma}.
\end{eqnarray}
The continuity equation is
\begin{equation}
\label{sigma_evol}
\frac{\partial \Sigma}{\partial t}
+
\frac{1}{r} \frac{\partial}{\partial r} ( \Sigma r v_{\mathrm{r}} )
+
\frac{1}{r^{2}} \frac{\partial}{\partial \phi} (\Sigma r v_{\phi})
=
0.
\end{equation}
A similar equation holds for the vertical component of the 
magnetic field in the limit of complete flux freezing:
\begin{equation}
\label{bz_evol}
\frac{\partial B_{\mathrm{z}}}{\partial t}
+
\frac{1}{r} \frac{\partial}{\partial r} ( B_{\mathrm{z}} r v_{\mathrm{r}} )
+
\frac{1}{r^{2}} \frac{\partial}{\partial \phi} (B_{\mathrm{z}} r v_{\phi} )
=
0, \label{indu}
\end{equation}
which expresses that the magnetic flux density $B_z$ is conserved.

The gas pressure is computed from the vertically integrated 
internal energy $e$ assuming an ideal gas for the equation of state, 
$P=(\gamma-1)e$, with ratio of specific heats $\gamma$. For the
computations reported here, the value $\gamma=1.4$ was used.
The adiabatic evolution of the internal energy is given by
\begin{eqnarray}
\label{e_evol}
\frac {\partial{ e}}{\partial t} 
&+& 
\frac{1}{r} \frac{\partial}{\partial r} (r e v_{\mathrm{r}})
+
\frac{1}{r^{2}} \frac{\partial}{\partial \phi} (r e v_{\phi}) 
\\ \nonumber
&=&
-
P \left[\frac{1}{r} \frac{\partial}{\partial r} (r v_{\mathrm{r}}) 
+
\frac{\partial}{r \partial \phi} (v_{\phi}) \right].
\end{eqnarray}

To close our set of equations we have to determine the inclination 
of the magnetic field lines to the surface of the accretion disc,
i.e. we need to derive $B_{\mathrm{r}}^{+}(r, \phi)$ and $B_{\phi}^{+}(r, \phi)$. 
Their values determine the magnetic forces in the equations of motion.
By our assumption of a potential field in the magnetosphere, 
electric currents only exist within the plane of the 
accretion disc, and the vertical component of the electric current 
vanishes everywhere.

The current is therefore of the form
\begin{equation}
\label{delta_j}
{\bf j}_{\mathrm{h}}(r,\phi,z)={\bf j}_{\mathrm{h}}(r,\phi) \delta(z) 
\quad \mbox{and} \quad  j_{\mathrm{z}}(r,\phi,z)=0
\end{equation}
where $\delta(z)$ is the Dirac--delta function.
A subscript $_{\mathrm{h}}$ denotes vectors 
parallel to the plane of the accretion disc.
Disk winds and ionized particles in the disc magnetosphere, 
neglected in our model, will certainly contribute to current 
flows in the disc magnetosphere. 
The calculation of the 3--dimensional disc magnetosphere 
in force-free approximation is, however, beyond the present 
computational feasibility.
The assumption of a potential field for the structure 
of the disc magnetosphere is equivalent to assuming 
that the magnetosphere is sufficiently close to its minimum
energy state.

In the thin disc limit, the magnetic field at the disc surface 
$(B_{\mathrm{r}}^{+}, B_{\phi}^{+}, B_{\mathrm{z}})$ 
is connected to the disc currents by
\begin{equation}
\label{connection}
B_{\mathrm{r}}^{+} = \frac{2 \pi}{c} j_{\phi},
\qquad
B_{\mathrm{\phi}}^{+} = - \frac{2 \pi}{c} j_{\mathrm{r}}
\end{equation}
and 
\begin{eqnarray}
\label{Bz_j}
B_{\mathrm{z}}(r,\phi) &=& 
\frac{1}{c} \int_{0}^{2\pi} \int_{r_{\mathrm{in}}}^{r_{\mathrm{out}}} 
\frac{ \partial_{\mathrm{r^{\pr}}} ( r^{\pr} j_{\phi} ) - 
       \partial_{\phi^{\pr}} j_{\mathrm{r}}}
     {R} {\mathrm{d}} r^{\pr} {\mathrm{d}} \phi^{\pr}
\\ \nonumber
&-&
 \frac{1}{c}\int_{\delta F_{disc}} {1\over R} \bmath{j(r^{\pr})}\cdot{\mathrm{d}}\bmath{l^{\pr}}
\end{eqnarray}
where the second term on the right hand side (RHS) 
of eq. (\ref{Bz_j}) is the current  
flow at the disc--boundaries and $R$ is given by 
\begin{equation}
R^2=r^2 + r^{\pr 2} -2 r r^{\pr} \cos(\phi - \phi^{\pr}).
\end{equation}

Eq. (\ref{Bz_j}) is derived from a partial integration of 
the vector potential field $\bmath{A}$ in Coulomb gauge
so that ${\bmath B}=\bgrad \times {\bmath A}$, and
(Landau \& Lifshitz 1975):
\begin{equation}
\label{Vec_A}
  \bmath{A}(r,\phi) = \frac{1}{c} \int_{0}^{2\pi} \int_{r_{\mathrm{in}}}^{r_{\mathrm{out}}} 
  \frac{\bmath{j}(r^{\pr},\phi^{\pr})}{R} {\mathrm{d}} r^{\pr}
{\mathrm{d}} \phi^{\pr}.
\end{equation}

Given the magnetic flux distribution in the disc $B_{\mathrm{z}}(r, \phi)$
eq. (\ref{Bz_j}) has to be inverted to yield the currents ${\bmath j}_{\mathrm{h}}$ 
in the disc. The inversion is unique by applying the fact that  currents do not accumulate,  ${\mathrm{div}}{\bmath j}=0$:
\begin{equation}
\label{currents}
\frac{1}{r} \frac{\partial}{\partial r}(r j_{\mathrm{r}})
+
\frac{1}{r} \frac{\partial}{\partial \phi} j_{\phi}
= 
0.
\end{equation}

This connects the radial $j_{\mathrm{r}}$ and azimuthal $j_{\phi}$ component of the current, and has to be solved together with the inversion of eq. (\ref{Bz_j}).

As our accretion disc has a central hole, the solution of the potential 
field problem with $B_{\mathrm{z}}(r,\phi)$ given in the disc
is no longer unique, since the domain space is multiply connected. 
To any solution of the inversion problem (\ref{Bz_j}), an arbitrary multiple of the solution for $B_z=0$ can be added. This special solution consists of closed field lines wrapping through the hole of the disc and around its outer edge without crossing the disc itself (like the windings of a ring-core coil). 
We thus have additionally to specify the number of field lines 
which pass through the central hole, i.e. the total magnetic flux
through the hole of the disc (see also Lubow, Papaloizou \& Pringle 1994)
\begin{equation}
\label{Psi}
\Psi=\int_{0}^{2\pi} \int_{0}^{\mathrm{r}_{\mathrm{in}}} r \, B_{\mathrm{z}}(r, \phi, z=0)
\, {\mathrm{d}} r \, {\mathrm{d}} \phi.
\end{equation}
A Fourier Transform of eq. (\ref{Psi}) shows that only the 
axisymmetric component contributes to the magnetic flux 
through the disc hole whereas all other components
cancel. The degree of freedom introduced by the presence of a hole thus enters only into the computation of the axisymmetric component of the field. 

Eqs. (\ref{sigma_evol}), (\ref{bz_evol}) and (\ref{e_evol})
specify the time evolution of $\Sigma$, $B_{\mathrm{z}}$ and $e$
when the velocities $v_{\mathrm{r}}$ and $v_{\phi}$ in the plane of the 
accretion disc are known. 
These follow from Euler's equation
(\ref{euler_1}) and (\ref{euler_2}) which includes thermal, gravitational 
and magnetic forces. 
To solve for the magnetic forces we invert 
equation (\ref{Bz_j}) with the differential constraint (\ref{currents})
for the unknown currents, and an assumed value for the magnetic flux though
the hole. 
Eq. (\ref{connection}) gives the relation 
between $B_{\mathrm{r}}^{+}$, $B_{\phi}^{+}$ and $j_{\phi}$, $j_{\mathrm{r}}$
and thus the magnetic forces are determined.
In the next chapter we show how we solve the hydrodynamic equations
and the magnetospheric potential problem numerically.

\section{Numerical Method}

We solve the equations on an Eulerian grid with equidistant 
spacing in the $r$ and $\phi$ 
coordinates. The inner disc rim is 
at $r_{\mathrm{in}} = 0.1 \, r_{\mathrm{out}}$ with $r_{\mathrm{out}}$
the radius of the outer edge of the disc. 
The number of grid points and grid spacing in radial direction are $n_r$ 
and $\Delta r = (r_{\mathrm{out}} - r_{\mathrm{in}})/n_{\mathrm{r}}$.
We use a staggered grid such that 
the scalar quantities $\Sigma, B_{\mathrm{z}}, e$
are defined at the 
cell centres and the vector quantities
$(v_{\mathrm{r}}, v_{\phi})$, $(B_{\mathrm{r}}, B_{\phi})$ and 
$(j_{\phi}, j_{\mathrm{r}})$ at the cell boundaries.
The equations are used in dimensionless 
form, as follows. As unit of length we take the outer disc radius, 
as the unit of time the inverse 
of the Keplerian rotation rate at the outer edge. 
Thus, the Keplerian velocity at the outer edge 
is unity, and the time for one Keplerian orbital 
period of the outer edge is $T=2\pi$.
For all the calculations presented here
we use $n_{\mathrm{r}} \times n_{\phi} =  156 \times 128 $
grid cells in $r$ and $\phi$ direction respectively.
We also performed some calculations on a smaller grid of
$64 \times 64$ grid cells. 
Comparing these models to calculations done with a higher resolution
we find only differences on the order of the applied grid spacing. 
We are thus convinced that the models presented here are resolved 
sufficiently.

\subsection{The hydrodynamic part}

The hydrodynamic part of the calculations is 
done with a natural extension to the scheme described in
Stehle \& Spruit (1999).

The equations are written in conservative form.
Terms in the equations are divided into advection 
and source terms.
The advection from one grid cell to the other is done
with the upwind--differencing scheme of van Leer (1977).
Thermal pressure, gravitational forces and compressional heating 
are calculated following Stone \& Norman (1992). 
Viscous processes and radiative cooling from the surface of the accretion 
disc are neglected.

The induction equation, i.e. eq. (\ref{bz_evol}),
is an additional equation compared 
with the hydrodynamic case. 
It has the same forms as the continuity equation, 
and is treated numerically in the same way.

The magnetic force is a new term in the equations of motion.
Contrary to the pressure force, which is calculated by 
a local derivative, the magnetic forces are given
by the bending of the magnetic fields where they pass through
the accretion disc.
The inclination of the magnetic field lines to the accretion disc
depends, however, on the global magnetic field structure
in the magnetosphere and only when this is known,  the magnetic 
forces be calculated.
We introduce the numerical method to calculate the magnetic forces
in Section (\ref{mag_solv}).

The time step is controlled by the Courant--Friedrich--Levy condition. 
In addition to the sound speed, there is a magnetic wave speed in the
problem. 
The wave is compressive, but since the restoring force results 
from the change in the external potential upon compression of the field
lines rather than of the magnetic pressure itself, 
it is a dispersive wave.
The phase velocity is (ST, Tagger et al. 1990)
\begin{equation}
\frac{\omega} {k} = \frac{B_{\mathrm{z}}} {\sqrt{2 \pi \Sigma k}}, 
\end{equation}
where $k=2\pi / \lambda$ is the wavenumber and $\lambda$ the
wavelength. 
The group speed, which carries the wave 
information, is a factor 2 lower.
The highest magnetic wave speed in the discretized 
equations is therefore obtained for the highest
wavenumber that can be represented by the grid. 
By the Nyquist theorem, this is $k=\pi/\Delta r$.
The magnetic wave speed $v_{\mathrm{m}}$ that enters 
the Courant condition is thus
\begin{equation}
v_{\mathrm{m}} = \frac{B_{\mathrm{z}}} {2 \pi \sqrt{2 \Sigma/\Delta r}}.
\end{equation}
A Courant factor of 0.75 was found to be sufficient 
for numerical stability.
In most cases, however, we find the time step to be controlled 
by the azimuthal velocity $v_{\phi}(r_{\mathrm{in}})$
at the inner accretion disc rim.
The additional magnetic wave  constrains
the time step only for models where the magnetic acceleration
$g_{\mathrm{m}}$ approaches the acceleration of gravity $g$.

The boundary conditions used are solid boundaries at the inner and outer
edges of the disc, i.e. $v_r=\partial P/\partial r=0$. These are sufficient
for the present calculations in which only the short-term evolution of the
disc is followed. For the longer-term evolution, one would want to use conditions
that allow accretion to take place through the boundaries. This is beyond the
scope of the present study.

\subsection{The solver for the disc magnetosphere}

\label{mag_solv}

The magnetic forces are calculated at each 
time step from the magnetic flux distribution 
$B_{z}(r,\phi)$ in the disc. 
This involves two steps. 
First, eq. (\ref{Bz_j}) is inverted to 
obtain the currents $\mathbf{j}_{\mathrm{h}}$ from the flux
distribution $B_{z}$. 
In the second step, the forces $B_{z}B^{+}_{r}/2\pi$, 
$B_{z}B^+_\phi/2\pi$ are computed using eqs. 
(\ref{connection}), and added to the hydrodynamic 
forces.

The inversion of eq. (\ref{Bz_j}) has two complications. 
Because of charge conservation the current is not 
an arbitrary function of $r$ and $\phi$, but must 
satisfy eq. (\ref{currents}), i.e. 
$\mathrm{div} \, {\mathbf j}=0 $.
This condition can be used to eliminate one of the 
current components $j_r$ and  $j_\phi$ in favor of the other. 
Eq. (\ref{Bz_j}) can then be read as an integral 
equation determining one of the current components 
in terms of the vertical field component.

Second, the region on which the computations are 
done is not simply connected. 
The central hole in the computational domain
generates an additional parameter in the potential problem, 
namely the net magnetic flux through the hole $\Psi$.
For the present exploratory calculations, we use fixed boundary 
conditions. Hence $v_r=0$ at the boundaries, and with the 
induction equation (\ref{indu}) no magnetic flux enters or leaves
through the edges of the disc. This also implies that $\Psi$ is constant
in time. More general boundary conditions are possible that 
take account of the advection of magnetic flux into the hole.
These will be needed in more realistic 
calculations of discs accreting on to protostars, for example. 

\subsubsection{Fourier decomposition in $\phi$}

Since the equations for the potential problem are 
homogeneous in the azimuth $\phi$, Fourier transforms 
can be used in this direction. 
Since they are also linear, the Fourier components 
do not mix, and one can solve for each of the Fourier components
separately. 
If the number of azimuthal grid points or Fourier 
modes is $n_{\phi}$, this reduces the computing effort 
required by a factor of the order $n_{\phi}$ compared with
straightforward discretization in $\phi$. 
By using Fourier decomposition, the computing effort 
for the potential problem can be kept at a level comparable to the
hydrodynamic parts of the calculation.

Thus we write the magnetic flux distribution as
\begin{equation}
B_{\mathrm{z}}(r, \phi) = B_{\mathrm{z}}^{0} + 
\sum_{m=1}^{\mathrm{n}_{\phi}/2}
\left( B_{\mathrm{z}}^{m,\mathrm{s}} \sin (m\phi) +
       B_{\mathrm{z}}^{m,\mathrm{c}} \cos (m\phi) \right),
\end{equation}
where
$B_{\mathrm{z}}^{0}(r)$, $B_{\mathrm{z}}^{m,\mathrm{s}}(r)$ 
and  $B_{\mathrm{z}}^{m,\mathrm{c}}(r)$ are only functions of $r$.
Similar equations hold for the currents $j_r, \, j_\phi$.

The Fourier amplitudes are given by 
\begin{equation}
B_{\mathrm{z}}^{m,\mathrm{c}}(r)=\frac{1}{\pi}
\int_0^{2\pi}B_{\mathrm{z}}(r,\phi) 
\cos(m \phi) {\mathrm d}\phi,
\end{equation}
and similarly for $B_z^{m,\mathrm{s}}$.
Using (\ref{Bz_j}) this becomes
\begin{eqnarray}
 B_{\mathrm{z}}^{m,\mathrm{c}}(r) & = & \frac{1}{\pi c}
          \int_{r_{\mathrm in}}^{r_{\mathrm out}}
          \int_{0}^{2\pi}
          \int_{0}^{2\pi} 
          {\mathrm d}r^\p
          {\mathrm d}\phi 
          {\mathrm d}\phi^\p
\\ \nonumber
& \left. \right. &
          \cos(m\phi) 
          \frac{ \partial_{r^\p}
[r^\p j_\phi(r^\p,\phi^\p)]-\partial_{\phi^\p} 
                  j_r(r^\p,\phi^\p) }  
               { (r^2+{r^\p}^2-2rr^\p\cos(\phi-\phi^\p) )^{1/2} }.
\end{eqnarray}
This can be written as 
\begin{eqnarray}
B_{\mathrm{z}}^{m,c}(r) & = & \frac{1}{\pi c \, r}
         \int_{r_{\mathrm in}}^{{r_{\mathrm out}}} 
         \int_{0}^{2\pi}
         {\mathrm d}r^\p
         {\mathrm d}\phi^\p
\\ \nonumber
& \left. \right. &
         \cos(m\phi^\p)
         [\partial_{r^\p}(r^\p j_\phi)-\partial_{\phi^\p}j_r]K_{m}(r^\p/r),
\end{eqnarray}
where
\begin{equation}
K_{m}(x) = \int_{0}^{2\pi} 
  \frac{ \cos (m \phi)}{( 1+x^2-2x \cos(\phi^{\pr}))^{1/2}} 
  \mathrm{d} \phi^{\pr} .
\end{equation}
We evaluate this function numerically. For $m=0$ 
it
can be expressed 
in terms of 
the complete elliptical function of the second kind $F(x)$ (Gradstein 
\& Ryzhik 1981),
\begin{equation}
K_{0}(x)=4 \chi F(\chi) \quad \mbox{with} \quad \chi=\min(x, 1/x).
\end{equation}

Substituting the Fourier expansions of $j_{\mathrm{r}}$ and $j_\phi$ and 
integrating over
$\phi^\p$, we get
\begin{equation} 
B_{z}^{m,c}(r)=\frac{1}{cr}
         \int_{r_{\mathrm in}}^{{r_{\mathrm out}}}
         [\partial_{r^\p}(r^\p j_\phi^{m,\mathrm{c}}) - m \, 
         j_{r}^{m,\mathrm{s}}]K_{m}(r^\p/r)
         {\mathrm d}r^\p.
\end{equation}
For $m \ne 0$ the charge conservation condition 
(\ref{currents}) 
can be used to eliminate $j_\phi$, which yields
\begin{eqnarray}       
\label{mnot0}
B_{z}^{m,c}(r) &=&  {\cal I}^{m,c} (r, j_{r}) = 
\frac{1}{c \, r}
\int_{r_{\mathrm in}}^{r_{\mathrm out}} 
{\mathrm d}r^\p 
\\ \nonumber
  &\left. \right.& 
      \left[ \frac{1}{m} 
             \partial_{r^\p}({r^\p}\partial_{r^\p} {r^\p}  
             j_{r}^{m,s})
         -
             m \partial_{\phi^\p}
             j_{r}^{m,s} 
     \right] K_m(r^\p/r),
\end{eqnarray}
where ${\cal I}^{m,\mathrm{c}} (r, j_{\mathrm{r}})$ is an abbreviation for the integral operator.
A similar equation holds for the $B_{\mathrm{z}}^{m,\mathrm{s}}$ component.

The axially symmetric component reads
\begin{equation} 
B_z^0(r)= {\cal I}^{0}(r, j_{\phi}) = 
         \frac{1}{c}
         \int_{r_{\mathrm in}}^{r_{\mathrm out}}
         \partial_{r^\p}({r^\p}j_\phi^0) K_0(r^\p/r) 
         {\mathrm d}r^\p.
  \label{mis0}
\end{equation}

The flux through the hole in the middle of the disc is
\begin{eqnarray} 
\label{ps0}
\Psi&=&{\cal I}_{\Psi}(j_{\phi}) = 
     2\pi 
     \int_0^{{r_{\mathrm in}}}rB_{\mathrm{z}}^0{\mathrm d}r  =
\\ \nonumber
     &=&
     \frac{2 \pi}{c}
     \int_{0}^{r_{\mathrm{in}}}
     \int_{r_{\mathrm in}}^{{r_{\mathrm out}}}
     \partial_{r^\p}({r^\p}j_\phi^0) K_{0}(r^\p/r) 
     {\mathrm d}r^\p
     {\mathrm d}r.
\end{eqnarray}

Eqs. (\ref{mnot0}) and (\ref{mis0}) are integral equations 
for the $(n_{\phi}-1)$ Fourier amplitudes  
$j_{r}^{m,\mathrm{c}}(r)$ and $j_{r}^{m,\mathrm{s}}(r)$
of the radial current, and the azimuthal current distribution 
$j_{\phi}^{0}(r)$. 
To solve these, we take finite differences in $r$, 
which turns each of the eqs. (\ref{mnot0}) and (\ref{mis0}) 
into a set of linear algebraic equations. 
The matrices involved are fixed in time and 
need to be inverted only once. 
If the number of grid points in $r$ is $n_{\mathrm{r}}$, 
the computing effort for the potential problem 
is therefore of the order $n_{\mathrm{r}}^2n_{\phi}$ 
per time step,
or $\sim n_{\mathrm{r}}$ per grid point and time step. 
Since the number of operations per grid point 
and time step in the hydrodynamic part of the 
calculation is a substantial, but fixed number 
independent of $n_{\mathrm{r}}$, the computing expense for 
the potential problem does not dominate the 
overall expense except for very large numbers 
of radial grid points.
In fact, for a grid of 
$n_{\mathrm{r}} \times n_{\phi} = 156 \times 128$ 
grid cells the magnetic part of the calculation 
takes about 50\% of the CPU--time.

\subsubsection{Discretization in $r$}

The discretization in $r$ of the integrals in 
(\ref{mnot0}) \& (\ref{mis0}) is different 
for the non--axisymmetric components (\ref{mnot0}) 
and the axisymmetric component (\ref{mis0}). 
The Fourier amplitudes $B_{\mathrm{z}}^m$ of the field are 
naturally defined at the same radial positions 
as the values of the field $B_{\mathrm{z}}$ itself, 
i.e. at the centres $r_i$ of the cells defined in the 
hydrodynamic part of the scheme. 
To discretize (\ref{mis0}), the currents $j_\phi^0$ 
are defined at the boundaries $r_{i+1/2}\equiv(r_i+r_{i+1})/2$ 
between these cells. 
Because of the topology of the domain used 
(a disc with a hole), there is one more of such 
boundaries than there are cell centres. 
Since the boundary conditions do not impose 
constraints on the azimuthal current at the boundaries, 
there is then also one more current $j^0_\phi$ than 
there are field amplitudes $B^0$. 
This additional degree of freedom is balanced by the the hole-flux 
condition (\ref{Psi}), which arises from the same 
topological property. 

\subsubsection{The axisymmetric component}

To evaluate the integral in (\ref{mis0}) we interpolate 
the current linearly between the values $J_j^0$ at 
its grid points:
\begin{equation}
j_{\phi,j}^{0}(r) = (x-1)J_j^0+xJ_{j+1}^0, 
\end{equation}
where $x=(r-r_{j+1/2})/(r_{j+3/2}-r_{j+1/2})$.
Inserting this into (\ref{mis0}), $B_z^0$ is a 
linear function of the currents $J_j^0$, 
with coefficients $B_{ij}^0$:
\begin{equation}
B_z^0(r_i)=\sum_j B_{ij}^0 \, J_j^0.
\end{equation}
Evaluation of these coefficients involves integrals of the type
\begin{equation}
\label{Int_axi}
   I^{0}(r)=\int_{r_{\mathrm{a}}}^{r_{\mathrm{b}}}
   ( \alpha + \beta r^{\pr} ) K_{0}(r^{\pr}/r_{i}) 
   {\mathrm{d}}r^{\pr}.
\end{equation}
The elliptic function involved in $K_0$ is 
evaluated by polynomial approximation
(Abramowitz and Stegun 1984), the integral 
by Bulirsch's algorithm (Press et al.\ 1995).

Similarly, assigning the index $i=0$ to the hole-flux 
$\Psi$, (\ref{ps0}) can be written as
\begin{equation}
\Psi=\sum_j B_{0j}^0 J^0_j \, .
\end{equation}
The coefficients $B_{ij}^0$ then form a square matrix of 
dimension $n_r+1$, relating a vector of magnetic variables, 
${\mathrm b}^0 = ( \Psi, B_{\mathrm{z}}^{0}(r_{1}), ... B_{\mathrm{z}}^{0}(r_{n_{r}}) )$, 
to the currents:
\begin{equation}
\label{matrix_0}
b_{i}^0 = \sum_{j=0}^{n_r} B_{ij}^0 J_{j}^0 \, .
\end{equation}

Since we use a fixed Eulerian grid, the matrix elements 
$B_{ij}^0$ are fixed and the inversion of the matrix can be done 
once and for all for a given computational grid.
This inversion is done by the LU decomposition method 
(Press et al. 1995).

\subsubsection{The non--axisymmetric components}

The procedure for the non--axisymmetric components is very similar to that 
for the axisymmetric component, except that $\Psi$ does not appear because it
is already determined by the axisymmetric component.
For $m \not =0$ the currents $J_{\mathrm{r}}^m$, like the 
$B_{\mathrm{z}}^m$, 
are defined at the cell centres, so that there
is an equal number of each.

Because we have used a partial integration in 
deriving (\ref{mnot0}), and have used charge 
conservation to eliminate $j_\phi$, the 
integrand in (\ref{mnot0}) contains a second 
derivative with respect to $r$. 
In order to evaluate it at the same order of 
accuracy as the axisymmetric coefficients, 
a third order interpolation is needed. 
We choose cubic spline interpolation
between neighbouring grid--cells. 
The coefficients then involve expressions of 
the type

\begin{equation}
{\cal K}_{m}^{l} = \int_{0}^{x} x^{l} K_m(x) 
\mathrm{d} x,
\end{equation}
with  $l=1,2,3$. 
Coefficients up to $m=64$ were evaluated with an accuracy of 
$10^{-7}$.

\subsubsection{Test calculations}

\label{mag_test}

\begin{figure}
  \centerline{\psfig{file=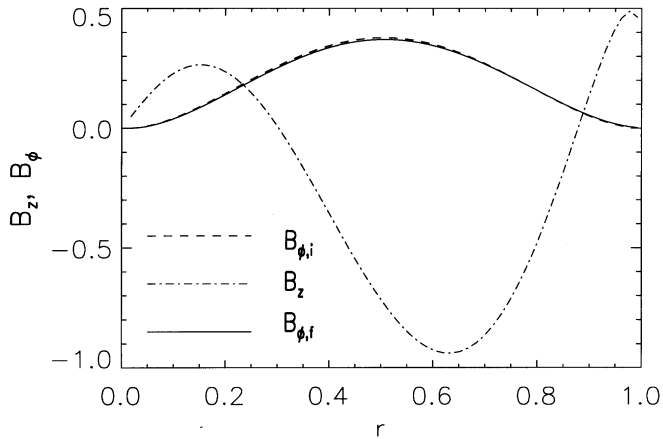,height=6.5cm,width=9.5cm}}
  \caption{
     An example showing the accuracy of the magnetospheric  
     field calculation for an $m=1$ mode. 
     The assumed azimuthal field at the top 
     surface of the disc $B_{\phi,\mathrm{i}}(r)$ (dashed) 
     agrees with the reconstructed values (solid)
     to 1 \% for this grid with 100 points in $r$.
  }
  \label{Fig_test}
\end{figure}  
To test the accuracy of our numerical solution of the 
potential field problem, we apply the inversion of 
eq. (\ref{Bz_j}) to a known field configuration. 
We proceed by first 
specifying ${\bmath j_{\mathrm{i}}}(r)$ analytically. 
Then we derive $B_{\mathrm{z}}(r)$ 
from a numerical integration of eq. (\ref{Bz_j}). 
The integration yields $B_{\mathrm{z}}(r_{\mathrm{i}})$
at the grid cell centres $r_{\mathrm{i}}$ with an accuracy of $\sim 10^{-6}$.
These values are used to invert $B_{\mathrm{z}}$ with our
magnetosphere--solver to reconstruct
the disc currents ${\bmath j}_{\mathrm{f}}(r_{\mathrm{i+1/2}})$ 
and field strengths which are then compared with the 
original function values at these points.

An example is shown in Fig. (\ref{Fig_test}).
This is an $m=1$ mode with the current distribution
\begin{equation}
j_{\mathrm{r,1}}^{\mathrm{m}=1} =  (r-r_{\mathrm{in}})^{2}(r_{\mathrm{out}}-r)^{2}
\end{equation}
with $r_{\mathrm{in}}=0.01 r_{\mathrm{out}}$.
The reconstructed current distribution matches the 
input distribution within an accuracy of 1\%, for the 100-point grid used.

Additional tests of accuracy of the code as a whole were done by
comparing results at different resolutions.

\section{Uniformly rotating discs}

The linear stability of uniformly rotating discs with magnetic fields 
of the type considered here was studied by ST. 
The condition for instability  of the interchange type is
\begin{equation}
\label{cond_inter}
a\equiv
\frac{B_{\mathrm{r}}^{+} B_{\mathrm{z}}}{2\pi\Sigma \Omega^{2}}
{\mathrm{d}\over\mathrm{d} r} \ln
\left( \frac{B_{\mathrm{z}}}{\Sigma} \right) < 0. \label{loc}
\end{equation}
Disks where $a \ge 0$ everywhere are stable to interchange instability. 
The derivation of this condition does not take account of 
possible global instabilities.

In this section we study two examples, 
one of a disc which is stable (model 1) 
and one that is  unstable (model 2) 
according to condition (\ref{loc}). 

Uniformly rotating discs are set up 
numerically by modifying the gravitational 
potential $\Phi(r)$ such that magnetic, 
centrifugal and gravitational forces are 
just balanced for the case $v_{\phi}(r)=$const. 
While this case is of limited astrophysical 
interest, it serves to test the agreement 
with the predictions from the linear theory, 
to check for possible global modes not 
covered by condition (\ref{loc}), and to 
get an impression of the nonlinear development 
of interchange instability in the present case.

We study the discs in a frame of reference 
which corotates with the disc. 
The corresponding non--inertial terms are 
added to the equations in section 2 
(see also Stehle \& Spruit 1999).
Both models presented here are advanced 
in time with zero magnetic flux through 
the central hole, i.e. with $\Psi=\mbox{const.}=0$.

The uniformly rotating case as defined above is 
a one-parameter family, governed by the ratio 
$c_{\mathrm{u}}$ of magnetic to the centrifugal forces:
\begin{equation}
c_{\mathrm{u}}=\left( \frac{g_{\mathrm{m}}}{g} \right)_{\mathrm{max}} = 
               \left( \frac{r B_{\mathrm{z}} B_{\mathrm{r}}}{2\pi \Sigma v_{\phi}^{2}} \right).
\end{equation}
Other parameters such as the amplitudes of the 
central potential, the magnetic field strength 
and the surface density can be scaled out of the 
equations.
In cases where $c_{\mathrm{u}} \ll 1 $ magnetic 
forces are unimportant and the disc rotates freely.
In the case $c_{\mathrm{u}} \to 1$ magnetic 
forces dominate and rotation can be neglected.

In the results shown, the gas pressure included in
the calculations for numerical reasons (section \ref{eqs}) has only a small
influence.

\subsection{Model 1: $\mathrm{d} 
            \left(B_{\mathrm{z}}/{\Sigma}\right)/{\mathrm{d}}r = 0 $}

\begin{figure}
  \centerline{\psfig{file=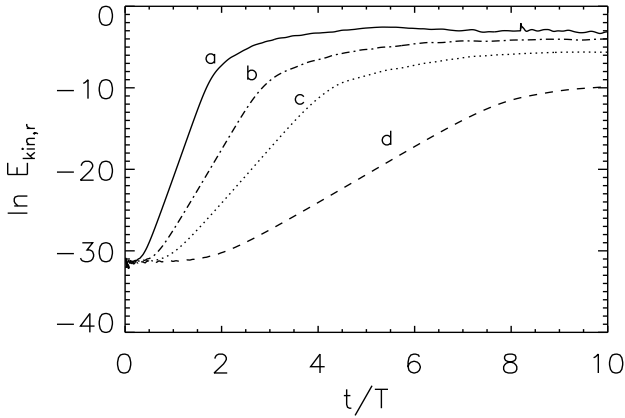,height=6.5cm,width=9.5cm}}
  \caption{The time evolution of the kinetic energy 
           $E_{\mathrm{kin,r}}$ for the uniformly 
           rotating model sequence. The duration 
           of one disc revolution at $r_{\mathrm{out}}$ is $2\pi T$.
The interchange instability grows on a dynamical time scale determined by the 
magnetic field strength. Both the growth rate and the saturation amplitude 
increase with the field strength.} 
 \label{Fig_uninot_Ek}
\end{figure}  

According to eq. (\ref{cond_inter}) discs with 
$\mathrm{d} \left(B_{\mathrm{z}}/{\Sigma}\right)/{\mathrm{d}}r = 0$
are expected to be stable against the interchange instability.
The initial density and magnetic field distributions 
are specified as Gaussian humps in $r$ 
[$\Sigma(r) \sim \exp( - (r-0.5 \, r_{\mathrm{out}})^2/\Delta^2 )$], 
with a maximum at $r=0.5 \, r_{\mathrm{out}}$ and width $\Delta =0.1 \, r_{\mathrm{out}}$.

Models with four different field strength are followed:  
a weakly magnetized disc ($ c_{\mathrm{u}} \simeq 9 \, 10^{-4}$), 
and one with a high magnetic support (model 1a, $ c_{\mathrm{u}}=225$).
Intermediate cases are chosen with $ c_{\mathrm{u}}=0.09$ and 
$ c_{\mathrm{u}}=9$ .
We perturb the initial stationary, 
axisymmetric models with a low amplitude 
($\sim 10^{-7} v_{\phi}(r_{\mathrm{out}})$)
point--to--point noise in $v_{\mathrm{r}}$.

We find the extreme cases $c_{\mathrm{u}} \to 0$ and 
$c_{\mathrm{u}} \to \infty$ to be stable.
The total kinetic energy in the radial velocity component.
(i.e. $E_{\mathrm{kin,r}} =  \int \Sigma v_{\mathrm{r}}^{2} {\mathrm{d}} F$)
was constant for the whole calculation of $\sim$15 disc orbits.

In the intermediate cases $c_{\mathrm{u}} \sim 1$ 
a very weak form instability was observed, 
with characteristics different from an interchange. 
The energy in the radial motions $E_{\mathrm{kin,r}}$ 
increased during the first 15 orbits by a factor 
$\sim 100$--$1000$ . 
A global disc pattern is excited, 
showing spiral arms with $m \sim 20$.
The wave saturates at a strength 
$(B_{\phi}/B_{\mathrm{z}}) \simeq 10^{-6}$--$10^{-4}$. 
The waves are of low amplitude and 
do not measurably transport angular momentum.
After a linear rise time of some orbits the disc is found to be 
stationary again, i.e. $\partial_{\mathrm{t}} \Sigma(r, \phi) = 0$,
even though it is now slightly non--axisymmetric.

The exact nature origin of this weak instability 
is not entirely clear at the moment.
In any case, the amplitude of the motions observed is 
low compared to those of the instabilities described below,
and are not relevant for actual accretion discs where the magnetic
support is always less than gravity, $c_{\mathrm u}<1$.

\subsection{Model 2:
 $\mathrm{d}\left(B_{\mathrm{z}}/{\Sigma}\right)/{\mathrm{d}}r < 0 $}

\begin{figure*}
  \caption{Interchange instability in a uniformly rotating disc with
  $B_z/\Sigma$ decreasing outward, and an initially uniform surface density $\Sigma$ (model 2b).
           The instability starts near to the inner edge 
           of the accretion disc, where the magnetic instability parameter 
           $a$ is largest, and spreads over the whole disc in a few orbits.} 
  \label{Fig_uninot_si}
\end{figure*}

Next we study accretion discs where condition 
(\ref{cond_inter}) predicts the interchange 
instability to be present.

We choose $\Sigma=$const. and $B_{\mathrm{z}}(r)$ 
decreasing with $r$ as a power law
\begin{equation}
B_{\mathrm{z}}(r)=B_{\mathrm{z},0} \left( \frac{r}{r_{\mathrm{out}}} \right)^{-4/5}.
\end{equation}
With this choice, the instability parameter $a$ has a
minimum at $r\approx 0.2 \, r_{\mathrm{out}}$. 
We expect the instability to first manifest itself near this radius.

\begin{table}
\caption[]{Parameters for model sequence 2: 
           initial values of the degree of instability $a_{\mathrm{m}}=\min(a)$ 
           and the ratio of magnetic to centrifugal acceleration 
           $c_{\mathrm{u}}$ at $r=0.2 \,  r_{\mathrm{out}}$, 
           and growth rate $\gamma_{\mathrm{E}}$ of the kinetic energy
           $E_{\mathrm{kin,r}}$.  }
\label{Pmag_Tab_2}
  \begin{center}
    \begin{tabular}{|c|c|c|c|}
       \noalign{\smallskip} 
       \hline
       \noalign{\smallskip}
        nr. &  $a_{\mathrm{m}}$ &    $c_{\mathrm{u}}$  &  $\gamma_{\mathrm{E}}$    
\\ 
\hline
   2a        &   $-1.10$   &      1.1               &   16.4
\\
   2b        &   $-0.36$   &      0.5               &    9.6
\\
   2c        &   $-0.16$   &      0.2               &    6.5
\\
   2d        &   $-0.05$   &      0.05              &    3.4   
\\
\hline
       \noalign{\smallskip}
    \end{tabular}
  \end{center}
\end{table}
As in model 1 we perturb the initial axisymmetric model with a low amplitude point--to--point noise in $v_{\mathrm{r}}$. 

Instability is found in all models of this sequence.
Fig. (\ref{Fig_uninot_Ek}) shows the time evolution of 
$E_{\mathrm{kin,r}}\equiv \int {1\over 2}\Sigma v_r^2 \, {\mathrm d}F $ 
for these models. The model parameters for this sequence are shown in 
Table \ref{Pmag_Tab_2}. 
At first $E_{\mathrm{kin,r}}$ increases exponentially.
$E_{\mathrm{kin,r}}$ saturates after several orbits and
the subsequent evolution is highly non--linear.

The instability causes a significant redistribution of the disc mass. 
This is illustrated in Fig. (\ref{Fig_uninot_si}), 
which shows the evolution of the surface mass density for model 2c.
It is seen that the instability first operates 
at $r \simeq 0.2 \, r_{\mathrm{out}}$ as expected from the local
minimum of $a$. 
A high mode number $m \simeq 15$ dominates at first. 
The pattern of motions resembles that of convective 
cells or the plumes of a Rayleigh-Taylor instability. 
The influence of rotation is evident in the asymmetry of the plumes.
With time the instability is seen to spread over 
the whole disc. The small instability cells merge and grow in size
as they drift to larger radii, as is characteristic of Rayleigh-Taylor instability. 

We identify this instability with the interchange 
instability as the disc pattern looks very similar 
to convective cells as predicted by Spruit et al. (1995),
and starts at the point where the linear 
analysis predicts the disc to be most unstable.

We conclude that the instability in the uniformly 
rotating case sets in as predicted from linear theory and 
has the nonlinear development of an interchange 
instability.

\section{Non--uniformly rotating discs}

We now study accretion discs revolving around the 
gravitational field of a point mass. 
According to the linear analysis of Spruit et al. (1995) 
and Lubow \& Spruit (1995), shear due to differential 
rotation acts as a stabilizing factor on the interchange 
instability. This analysis predicts that instability appears only in regions 
of the disc where magnetic forces contribute significantly 
to support against gravity. The predicted linear growth is algebraic
(a power law of time) rather than exponential.

We study the evolution of two different initial setups. 
First we follow a model where the magnetic field decreases 
as a power of radius  (model 3) and then a case
where it decreases exponentially (model 4).
The initial structure of the models is summarized 
in Tab. (\ref{Pmag_Tab_3}) and (\ref{Pmag_Tab_4}).
The initial $v_{\phi}(r)$ is found from the radial 
force balance between magnetic, gravitational 
and centrifugal forces. 
We then perturb $v_{\mathrm{r}}$ by point--to--point 
low amplitude noise and follow the subsequent disc evolution
numerically.
As before all models are calculated with zero 
magnetic flux through the central hole of the disc, 
 $\Psi=0$.

\subsection{Model 3: $B_{\mathrm{z}} \sim r^{-5/4}$ and $\Sigma \sim r^{-3/2}$}

\begin{figure}
  \centerline{\psfig{file=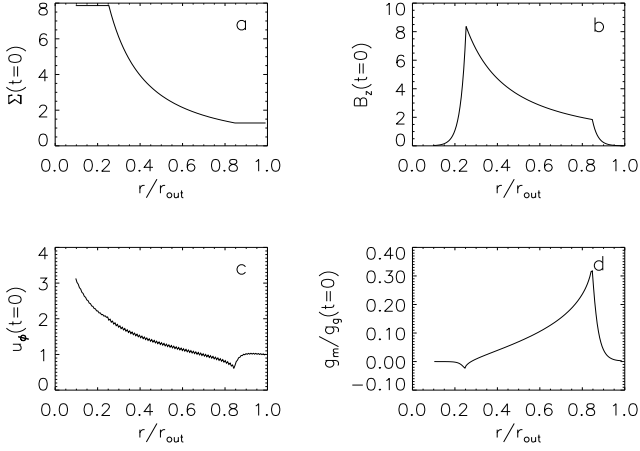,height=6.5cm,width=9.5cm}}
  \caption{The initial axisymmetric condition of model 3a. 
           Initial surface mass (a) magnetic field (b) 
           and azimuthal velocity distribution (c). The ratio of the
           magnetic ($g_{\mathrm{m}}$) and 
           gravitational ($g$) accelerations is shown in (d).}
  \label{Fig_M09_init}
\end{figure}

For the initial state in model sequence 3we choose a surface density 
varying as $\Sigma \sim r^{-3/2}$.
In order to contain the effects of the instability 
within the computational domain as much as possible, 
we choose a magnetic field distribution with strength 
vanishing towards the boundaries (Fig. \ref{Fig_M09_init}).

\begin{figure}
  \centerline{\psfig{file=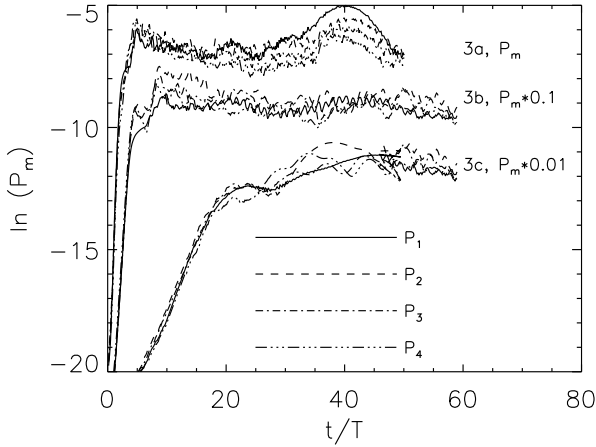,height=6.5cm,width=9.0cm}}
  \caption{Growth of the instability for model sequence,
          showing evolution of the power $P_m$ in the 
          first 4 Fourier modes, relative to the
          power in the axisymmetric mode. 
          Curves are shifted vertically by factors of 10
          since the saturation levels are nearly the same.
          The power in the 4 modes is comparable, except during the hump
          at $t/T \sim 40$ in model 3a (`outburst')
          when $m=1$ dominates.}
  \label{Fig_M09_mag}
\end{figure}

The degree of support against gravity by the magnetic field, 
as measured by the ratio $g_{\mathrm{m}}/g_{\mathrm{g}}$, 
increases with radius and has a maximum at 
$0.85 \, r_{\mathrm{out}}$ (Fig. \ref{Fig_M09_init}d).
The parameters of the model sequence are summarized in Tab. (\ref{Pmag_Tab_3}).
The parameter ${\cal R}$ specifies the mean magnetic flux per 
surface--mass:
\begin{equation}
\label{Pmag_calR}
   {\cal R}= \frac{ \int_{\mathrm{disc}} B_{\mathrm{z}} {\mathrm{d}} F }
               { \int_{\mathrm{disc}} \Sigma         {\mathrm{d}} F }.  
\end{equation}
and can be used to compare model calculations.

\begin{table}
\caption[]{Parameters for model sequence 3.
           The relative support of the disc by magnetic forces 
           $(g_{\mathrm{m}}/g_{\mathrm{g}})_{\mathrm{max}}$, the 
           initial growth rate 
           $\gamma_{\mathrm{Ekin,r}}$,
           the ratio ${\cal R}$ of the total magnetic flux through the disc
           to the total mass, and the mass accretion rates in units of 
           the total disc mass $M_{\mathrm{disc}}$ per time unit $T$. }
\label{Pmag_Tab_3}
  \begin{center}
    \begin{tabular}{|c|c|c|c|c|}
       \noalign{\smallskip} 
       \hline
       \noalign{\smallskip}
   model nr. &   $(g_{\mathrm{m}}/g)_{\mathrm{max}}$  &    $\gamma_{\mathrm{Ekin},r}$ 
 &
       ${\cal R}$    &   $\dot{M}(r < 0.3 r_{\mathrm{out}})$
\\ 
\hline
   3a ($t/T \le 35$)  &   0.32        &   27.1  &    0.93   &  
                             $2.0 \, 10^{-4} \, M_{\mathrm{disc}}/T $
\\
   3a ($t/T \sim 40$) &               &         &           &  
                             $1.7 \, 10^{-3} \, M_{\mathrm{disc}}/T $
\\
   3b        &   0.14        &   6.9   &    0.63   & 
                             $1.2 \, 10^{-5} M_{\mathrm{disc}}/T $
\\
   3c        &   0.035       &   0.96  &    0.31   &  
                             $\sim 4 \, 10^{-6} M_{\mathrm{disc}}/T $
\\
\\
\hline
       \noalign{\smallskip}
    \end{tabular}
  \end{center}
\end{table}

Instead of interchange type instability, we find, in all three cases 
that the initial set up is unstable to 
a global, non--axisymmetric instability. 
The wave pattern of the instability can be traced from one 
edge of the disc to the other (see Fig. \ref{Fig_M09_si}).
Initially the kinetic energy $E_{\mathrm{kin,r}}$ growth exponentially 
on a dynamical time-scale. 
The initial growth rate, as given by 
$\gamma_{\mathrm{Ekin,r}}={\mathrm d}\ln E_{\mathrm{kin},r}/{\mathrm{d}t} $,
is largest for the disc with the highest magnetic support.

Fig. (\ref{Fig_M09_mag}) shows the time evolution of the 
power 
\begin{equation}
P_{\mathrm{m}} = \frac
   {\int_{r_{\mathrm{in}}}^{r_{\mathrm{out}}}
  \left( (B_{z}^{m,\mathrm{c}})^{2} + (B_{z}^{m,\mathrm{s}})^{2} \right)^{1/2} \,
           r {\mathrm d} r }
   {\int_{r_{\mathrm{in}}}^{r_{\mathrm{out}}}
   B_{z}^{0} \,
           r {\mathrm d} r }
\end{equation}   
in Fourier mode $m$ of the field strength $B_{\mathrm{z}}$,  
integrated over the whole disc, and relative to the
axisymmetric component $m=0$. 
In all three calculations we find that the modes $m=1$--4 
grow equally fast.
The relative power saturates nearly at the same level, independent of the 
degree of magnetic support.
The relative power in the first 4 modes is similar.
For the calculation with the highest magnetic field strength, 
however, (model 3a) the $m=1$ component dominates for a period of
about 15 orbits around  $t/T \simeq 40$ (the numerical time unit $T=1/\Omega_{\mathrm{out}}$, where $\Omega_{\mathrm{out}}$ is the orbital frequency at the outer edge).

\begin{figure}
  \centerline{\psfig{file=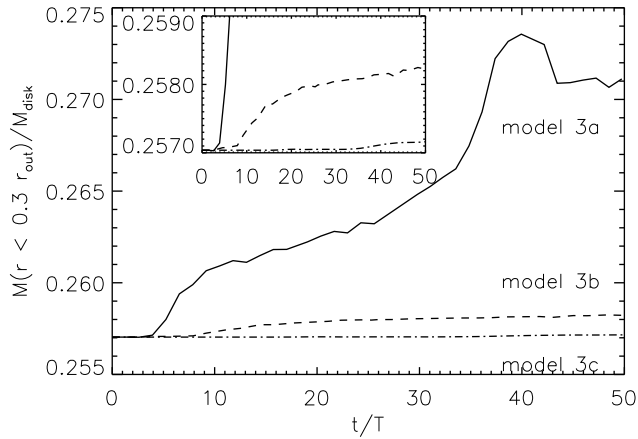,height=6.5cm,width=9.cm}}
  \caption{The evolution of the mass in the inner disc 
           ($r_{\mathrm{in}} \le r \le 0.3 \, r_{\mathrm{out}}$) 
           for models of sequence 3.
           Only model 3a shows a significant mass accretion towards 
           the central star. It shows a period of enhanced accretion around
           $t=40$ during which a prominent $m=1$ spiral arm is present,
           cf. Figures 5 and 7} 
  \label{Fig_M09_M}
\end{figure}  

Fig. (\ref{Fig_M09_M}) shows the evolution of the mass 
in the inner disc ($r_{\mathrm{in}} \le r \le 0.3 \, r_{\mathrm{out}}$)
in units of the total disc mass $M_{\mathrm{disc}}$.
Mass piles up in the inner part of the disc,
the faster the higher the magnetic support. 
This is accompanied by 
an outward transport of angular momentum by the magnetic instability.
For model 3a we find a roughly linear increase of the inner disc mass with time
corresponding to an accretion rate 
$\dot{M}_{\mathrm{disc}}(r<0.3r_{\mathrm{out}}) = 2 \; 10^{-4} M_{\mathrm{disc}}/T$,
but superposed on this trend is a much more active `outburst' around 
$t \simeq 40$.
During this active episode the accretion rate is about 10 times higher
(i.e. for $35 < t/T < 40$).
Fig. (\ref{Fig_M09_mag}) shows that during the outburst 
$P_{1}$ is larger by at least a factor 3--5 compared to the other modes, 
and by a factor $\sim10$ larger than in the preceding phase. 
Figs. (\ref{Fig_M09_si}) show that a strong m=1 spiral wave,
traveling outward from a crescent-shaped disturbance in the inner disc, is present during the outburst.

The time-scale for mass accretion in model 3a is longer, by a factor 
$10^{3}$--$10^{4}$, than the dynamical time-scale. 
For discs with less magnetic support the mass accretion time-scales are
so long that we have been unable to follow their evolution 
beyond the initial development of the instabilities.

\begin{figure*}
  \caption{Snapshots of the evolution of model 3a. Instability 
sets in with rather high $m$.
Later the modes $m=$1--4 are strongest until 
at $t/T \sim 35$ the $m=1$ component becomes dominant.
During the presence of the $m=1$ mode the mass accretion 
toward the central object is enhanced. The outward traveling 
$m=1$ spiral is generated by a rotating crescent-shaped 
enhancement which drifts inward with time.
           } 
  \label{Fig_M09_si}
\end{figure*}  

The evolution of the disc pattern as seen 
in $B_{\mathrm{z}}$ is shown in Fig. (\ref{Fig_M09_si}). 
The corresponding images of the surface density $\Sigma$ 
are found in Stehle (1997).
It is seen that the global instability starts with rather 
high mode numbers, $m \simeq 5$...8
(Fig. \ref{Fig_M09_si}a). 
Subsequently the waves are wound up (Fig. \ref{Fig_M09_si}b)
and it is only later that the modes $m=1$...4 become dominant.
After about 30 orbits of the outer disc edge, 
a prominent $m=1$ spiral arm develops. It
causes mass accretion rates 10 times higher than during the 
preceding phase.
The relative strength in the $m=$1 component 
is largest near the inner disc edge.
At $r \simeq 0.2 \, r_{\mathrm{out}}$ a prominent 
crescent-shaped field strength enhancement
shows up (best seen in Fig. \ref{Fig_M09_si}c). 
It is accompanied by a similar enhancement in 
the surface density (Stehle 1997). 
The crescent rotates approximately with the 
local orbital rate. 
An $m=1$ wave travels outward from this rotating crescent.
The  maximum of the crescent moves in to smaller 
radii (compare Figs. \ref{Fig_M09_si}c and e), where the 
rotation rates are higher. 
The outward traveling wave becomes correspondingly more tightly wound.

This behavior is very reminiscent of a purely hydrodynamic 
form of global instability observed in hot, partially 
pressure-supported discs 
(Blaes and Hawley 1988; R\'o\.zyczka \& Spruit 1993). 
This instability takes place only when
the degree of support against gravity by pressure 
becomes noticeable, and 
it also takes the form of a crescent rotating at 
the local orbital rate. 
It generates shock waves traveling outwards
and inwards. 
These waves cause the mass in the disc to spread, 
while the angular momentum lost by the crescent causes 
it to spiral in towards the centre, behaving
much like a solid object in doing so. 
Though the waves in the present calculation are 
rather different from hydrodynamic shock waves, 
we suspect that the same mechanism is at work. 
The peculiar behavior of the crescent mode and the fact that it appears
only at certain phases suggests that
it is a basically nonlinear phenomenon, not 
related directly to the linear
global modes of the system. The nature of this phenomenon
warrants further study.

\begin{figure*}
  \caption{The distribution of $B_{r}/B_{z}$ and $\ln \Sigma$ at
           the end of model calculation 3a ($t/T=62.4$). The surface 
           density decreases exponentially and the inclination of the 
           magnetic field lines towards the outer disc edge decreases.}
  \label{Fig_M09_final}
\end{figure*}

With time, mass piles up near to the inner and 
outer edge of the accretion disc.
The effect of the instability is thus much like that 
of viscous spreading, but it must be stressed that 
this is due to a global transport of angular momentum 
by the spiral wave, which can not be reduced to the 
action of a local viscosity.
After the approximately 10 outer disc orbits during which the $m=1$
component dominated, its relative strength compared to the 
other modes decreases again and the phase of high mass accretion rates 
is finished (Fig. \ref{Fig_M09_si}f).

At the end of the outburst the disc density distribution 
has completely changed. 
It now decreases approximately exponentially with radius 
rather than as a power law.
The same is true for the $B_{\mathrm{z}}$ distribution
(Fig. \ref{Fig_M09_final}). 

The dissipation taking place during the redistribution of
mass and magnetic flux causes the disc to heat up, so that
the gas pressure increases over its initial low value. At the
end of the calculation the Mach number of the orbital motion
in the inner disc has decreased to values of 5--10.

For the calculations done with magnetic support 
less than in model 3a the mass accretion rates 
are too low for significant redistribution of mass to occur
over the 50 orbits we were able to follow. 
It is thus unclear if discs with less magnetic support
also show outbursts, or if a threshold in the magnetic 
field strength exists below which the outburst mechanism 
cannot operate.

Since the outburst was a transient but obviously very 
effective transporter of mass and angular momentum, 
one wonders what caused its  decline. 
After the outburst the  degree of support of the disc 
against gravity by the magnetic forces has decreased substantially. 
Since the amplitude of all global waves seen in our results 
increases sharply 
with the degree of magnetic support, it is possible that 
the outburst declined because the crescent instability 
operates only at sufficiently high degrees of support.
This view is consistent with the properties of the 
purely hydrodynamic crescent instability. 

Another possibility is that an exponential dependence 
on radius is perhaps a more stable configuration, 
towards which the instability tends to develop. 
To test this possibility, we investigate in the 
next model a sequence of discs where $\Sigma$ and 
$B_{\mathrm{z}}$ decrease exponentially with radius.

\subsection{Model 4: $B_{\mathrm{z}}\sim\exp(-r)$ and $\Sigma \sim
            \exp (-r)$ } 

\begin{table}
\caption[]{As Tab. 3, but for model sequence 4} 
\label{Pmag_Tab_4}
  \begin{center}
    \begin{tabular}{|c|c|c|c|c|}
       \noalign{\smallskip} 
       \hline
       \noalign{\smallskip}
   model nr. &   $(g_{\mathrm{m}}/g_{\mathrm{g}})_{\mathrm{max}}$  &    $\gamma_{\mathrm{Ekin,r}}$ 
&
       ${\cal R}$    &   $\dot{M}(r < 0.3 r_{\mathrm{out}})$
\\ 
\hline
   4a        &   0.36        &   5.8    &   3.0     &   $8.7 \, 10^{-6}$M/T
\\
   4b        &   0.16        &   3.7    &   2.0     &   $6.2 \, 10^{-6}$M/T
\\
\hline
       \noalign{\smallskip}
    \end{tabular}
  \end{center}
\end{table}

\begin{figure}
  \centerline{\psfig{file=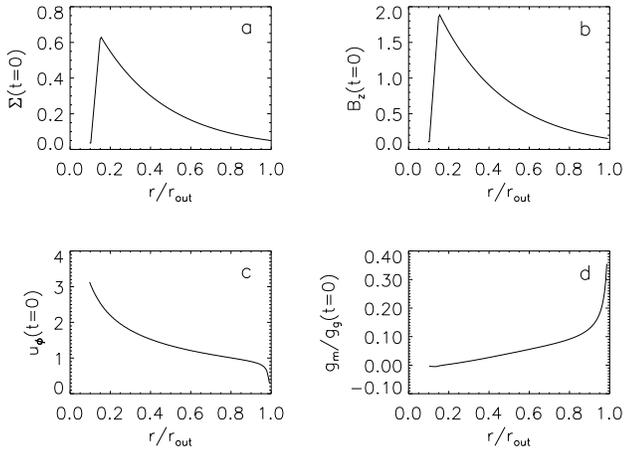,height=6.5cm,width=9.cm}}
  \caption{The initial surface density (a) magnetic field (b) and azimuthal 
           velocity distribution (c) of model 4a. 
           $B_{\mathrm{z}}(r) \sim \Sigma(r) \sim \exp(-r)$.  Panel (d)
           shows the initial ratio of magnetic to gravitational acceleration in
           the model. See text for further details. } 
  \label{Fig_M13_init}
\end{figure}

The final disc structure of model 3a motivates us to study 
the stability of accretion discs where the initial magnetic 
field and surface mass density decrease exponentially with radius.
In this sequence we study two such cases, which differ only in 
the initial field strength. 
In model 4a we choose a field with magnetic support
$(g_{\mathrm{m}}/g)_{\mathrm{max}}=0.36$ 
at maximum and in model 4b one with 
$(g_{\mathrm{m}}/g)_{\mathrm{max}}=0.16$.
The initial structure of model 4a is shown 
in Fig. (\ref{Fig_M13_init}).

We take $B_{z}/\Sigma=$ const. for the whole disc. 
$\Sigma(r)$ and  $B_{z}(r)$ decline linearly to zero at the inner edge.
The initial axisymmetric distribution is again perturbed 
in $v_{r}$ by a low amplitude point--to--point noise.

Values for the ratio ${\cal R}$
are listed in  Tab. (\ref{Pmag_Tab_3}) for models 3 and in 
Tab. (\ref{Pmag_Tab_4}) for models 4. 
A comparison of these values shows that the
total magnetic flux through the disc in models 4 is 
higher than in models 3, for the same degree of 
magnetic support. 
This difference is partly caused by the rapid decline 
of the magnetic field strength towards the disc edges 
in models 3 and partly from the fact that 
$\mathrm{d}(B_{\mathrm{z}}/\Sigma)/\mathrm{d} r > 0$ 
for the initial distribution in model 3.

\begin{figure}
  \centerline{\psfig{file=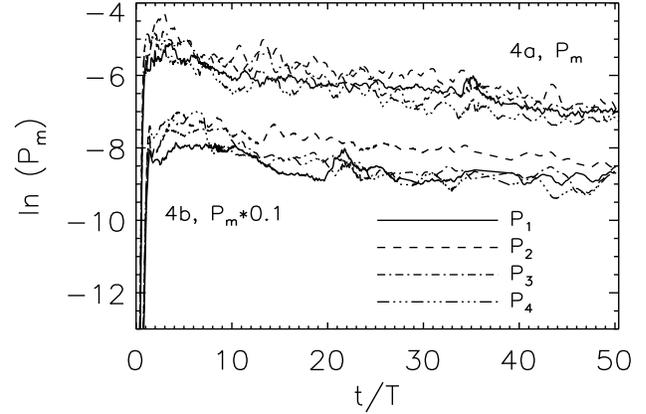,height=6.5cm,width=9.5cm}}
  \caption{The relative power $P_{\mathrm{m}}(t/T)$ for the modes $m=$1--4
           in models 4. 
           The values for model 4b are vertically shifted by multiplying 
           $P_{\mathrm{m}}$ with 0.1.
           $P_{\mathrm{m}=2}$ is most dominant and the $P_{\mathrm{m}=1}$
           component is weaker during most time of the integration.}
  \label{Fig_M13_Pm}
\end{figure}

Fig. (\ref{Fig_M13_Pm})  shows the time evolution of 
the relative power $P_{\mathrm{m}}$ in the Fourier 
modes $m=$1--4,  for models 4a and 4b. 
In both models the relative power $P_{\mathrm{m}}$ 
again increases initially on a short, dynamical, 
time-scale and the mode strength at which they
saturate is similar to what we observe in model sequence 3. 
As expected from model 3a at times after the outburst, 
$P_{\mathrm{m}=1}$ is comparable to $P_{\mathrm{m}=3}$ 
and $P_{\mathrm{m}=4}$ but significantly weaker than 
$P_{\mathrm{m}=2}$.
The $m=2$ mode appears to be the dominant mode 
during most time of our integration.

\begin{figure}
  \centerline{\psfig{file=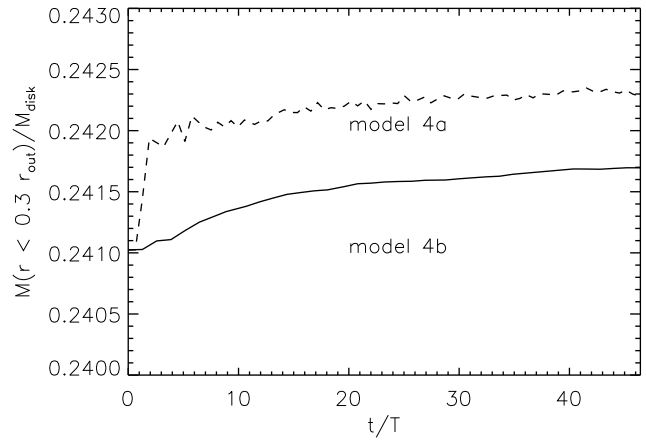,height=6.5cm,width=9.cm}}
  \caption{The evolution of the mass in the inner disc 
           $M(r<0.3 \, r_{\mathrm{out}})$ in models 4. 
           The accretion time-scales are large compared to models 3.} 
  \label{Fig_M13_M}
\end{figure}

Even though ${\cal R}$ is significantly higher than 
in model 3, the mass accretion rates at the inner part 
of the disc are small compared to model 3. 
The time-scale to clear the disc mass completely 
at these rates is now about $10^{5}$ outer edge orbits. 
This can be seen in Fig. (\ref{Fig_M13_M}), where we plot the 
evolution of the mass in the innermost part of the disc, 
$M(r<0.3 \, r_{\mathrm{out}})$ 
in units of the total disc mass $M_{\mathrm{disc}}$
(see also Tab. \ref{Pmag_Tab_4}).
It is also interesting to note that the 
mass accretion is now much less dependent on the strength 
of the magnetic field.
The low values for $\dot{M}(r<0.3 \, r_{\mathrm{out}})$ 
are accompanied by a low level of power in the $m=1$ mode.
This agrees with model 3, where the highest mass 
accretion rates are also found at a time where 
$P_{\mathrm{m}=1}$ is large.

\begin{figure}
  \caption{Radial velocity amplitudes
           at the end of model calculation 4a ($t/T= 51.1$). 
           Unit of velocity is $v_{\phi,\mathrm{Kepler}}(r_{\mathrm{out}})$. 
           The tightly wound spiral arms can be traced from one edge of the 
           disc to the other.}
  \label{Fig_M13b_pat}
\end{figure}

\begin{figure}
  \centerline{\psfig{file=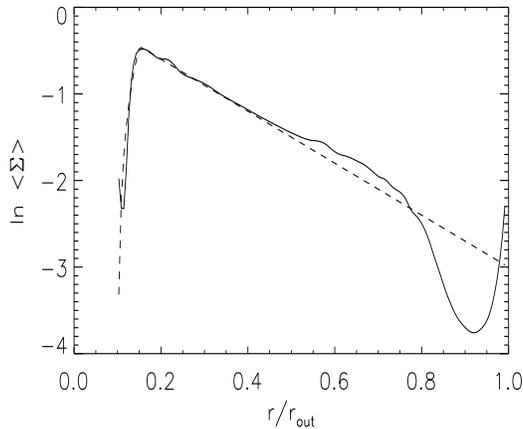,height=6.5cm,width=7.cm}}
  \caption{The azimuthally averaged surface density $<\Sigma(r)>$ 
           at the end of the calculation $t/T=51.1$ 
           (full line), compared with the initial distribution (dashed
           line). 
           Significant redistribution of mass has taken place only 
           in the outer parts of the disc.}
  \label{Fig_M13b_si}
\end{figure}  

In Fig. (\ref{Fig_M13b_pat}) we show the 
radial velocities of model 4a at $t/T=51.1$, the time where we 
stopped the integration, Fig. (\ref{Fig_M13b_si}) shows
the azimuthally averaged surface density $<\Sigma>$
at that time.

The pattern shows tightly wound spiral arms which 
can be traced from one edge of the accretion disc 
to the other. 
The radial velocities are rather small,
of the order $10^{-2}$ of the orbital velocity at 
the outer edge of the disc.
The mass redistribution has been strongest for 
$r \ga 0.7 \, r_{\mathrm{out}}$ where it 
has taken only a few binary orbits. This is just the region where
the magnetic support was strongest. In comparison, the mass 
redistribution in the inner regions is small. 

In both sequences 3 and 4 the redistribution of mass and magnetic flux
appears to be closely related to the degree of support against gravity by the
magnetic field configuration. It appears in regions where $g_{\mathrm{m}}/g$ exceeds
5--10 per cent. The comparison also shows that, unlike interchange instability, the global instability that causes this redistribution is not directly related to the 
flux-to-mass distribution $B/\Sigma$.

\section{Conclusions and discussion}

We have studied, by numerical simulation,  
the stability of accretion discs threaded by strong large scale magnetic fields,
assuming that the discs are geometrically thin. The simulations
solve the MHD equations for a vertically averaged accretion disc in
the $(r,\phi)$--plane.
The disc magnetosphere is calculated 
in potential field approximation, i.e. we treat the disc as a current sheet 
and assume the magnetic field outside the disc to be current free.
This approximation allows us to follow the evolution of a 3-dimensional
field configuration with the computational effort of a 2-dimensional simulation.

Previous analytical studies by Spruit \& Taam (1990), predicted that
uniformly rotating discs would show an interchange-type instability, a local
instability that appears
at any location where the magnetic field $B_z$ decreases with radius more rapidly
than the surface density $\Sigma$. To test this prediction, as well as the stability
of the numerical method, we first computed discs with initially uniform rotation.
The results agree with the analytic stability condition and growth rates. No 
additional, non--local, forms of instability were found in the simulations of
initially uniformly rotating discs. 
The nonlinear development of the instability agrees with 
that expected of an interchange instability like convection or Rayleigh-Taylor.

Linear analysis (Spruit et al. 1995; Lubow \& Spruit 1995) predicts 
that differentially rotating discs with approximately Keplerian rotation 
are much more stable than uniformly rotating discs. It predicts that 
instability of the interchange type occurs only when the local shear
 rate is less than the growth rate of the instability in a uniformly
 rotating but otherwise identical disc. For smooth distributions of
 $B_z$ and $\Sigma$ with $r$ this is equivalent to the condition
 that the magnetic forces contribute significantly to the support 
of the disc against gravity (Spruit et al. 1995). Such discs are strongly 
magnetized in the sense that $v_{\mathrm A}\gg \Omega H$, where
$v_{\mathrm A}$ is the Alf\'en speed at the midplane of the disc and
$H$ the disc thickness. The importance
of this prediction is that it suggests that even quite strong 
poloidal magnetic fields might still be stable in an accretion disc. 

To see if this prediction holds up in a full numerical simulation, we have done 
a sequence of calculations for discs with approximately Keplerian rotation, 
in which the magnetic contribution to support against gravity ranges from a few \% to 36\%.
At these values, the linear results predict a weak form of interchange instability,
with perturbations slowly growing as a power of $t$. 
In contrast with
the uniformly rotating case, however, no evidence of this form of
instability was found in the simulations.
 Instead, a new, global, exponentially growing form of instability appears.
It takes the form of tightly wound spiral arms which can be traced from one edge 
of the accretion disc to the other. 
This instability was found both in cases where the
magnetic field and the surface mass decrease with radius as a power
law (model sequence 3) and where they decrease exponentially (model
sequence 4). 
Its presence seems to depend primarily on the degree of magnetic
support of the disc; 
its amplitude is a steep function of this quantity.

The instability acts 
similar to viscous spreading in the sense that an outward 
angular momentum transport takes place which allows mass accretion from
 the outer to the inner regions of the disc. 
Mass accretion is of the order of $10^{-6}$--$10^{-4} \, M_{\mathrm{disc}}/T$ where 
$M_{\mathrm{disc}}$ is the 
total disc mass and T the Kepler time-scale $1/\Omega$ at the outer edge of the accretion disc.

In the case with the highest degree of magnetic support an additional, more
violent form of instability is observed. It takes place during a limited period of
$\sim 10$ disc orbits. The mass accretion rate in this episode is 10 times higher than
the long term average in the simulation. 
During this time the disc is perturbed by a strong one-armed spiral wave excited by a
density enhancement rotating at the local orbital frequency at the location of highest magnetic support. At the end of the outburst the surface density and the 
magnetic field distribution in the disc decline exponentially with disc radius 
and the subsequent mass accretion is less efficient.

The time-scale for the redistribution 
can be rather large compared to the dynamical time-scale.
Especially for discs with low magnetic support this time-scale 
is so large that, extrapolating the results from the simulation,
the final configuration would be obtained only after some 1000--100000 disc orbits.
Only when the magnetic support of the disc is of the order of the gravitational 
or centrifugal forces, does the time-scale become small enough that the  
redistribution of the disc mass proceeds on time-scales of several 10 disc orbits.

With the present results we can not establish whether the instabilities found
can also lead to instability and redistribution of mass and angular momentum at
lower degrees of magnetic support, or is limited to stronger fields. The difference between these possibilities is of obvious importance for accretion on longer 
time-scales. If the instabilities found here generally require magnetic support exceeding 
a few per cent, quite strong poloidal fields might exist in the inner regions of accretion discs.

In the protostellar context, such strong fields may also be relevant for the problem
of binary formation (cf. Sigalotti and Klapp 2000)

\bigskip

{\bf ACKNOWLEDGMENTS}

The work reported here was done as part of the research network 'accretion on to compact objects and 
protostars' supported by EC grant ERB--CHRX--CT93--0329. R.S. has been supported by 
the 'Deutscher Akademischer Austauschdienst im Rahmen des 2.Hochschulsonderprogramms'.  R.S. wants to thank J.Papaloizou for many stimulating discussions.

\bsp 

\label{lastpage}

\end{document}